\documentclass[runningheads]{llncs}
\usepackage[T1]{fontenc}
\usepackage{graphicx}
\usepackage{comment}
\usepackage{xcolor}
\usepackage{graphics}
\usepackage{graphicx}
\usepackage[font=bf]{caption}
\usepackage{hhline}
\usepackage{bigstrut}

\usepackage{colortbl}
\usepackage{graphicx}
\begin{document}
\title{The RPM3D project: 3D Kinematics for Remote Patient Monitoring\thanks{Supported by the ATTRACT project funded by the EC under Grant Agreement 777222.}}
%
%
\author{Alicia Fornés\inst{1}\orcidID{0000-0002-9692-5336} \and
Asma Bensalah\inst{1}\orcidID{0000-0002-2405-9811} \and
Cristina Carmona-Duarte\inst{2}\orcidID{0000-0002-4441-6652} \and
Jialuo Chen\inst{1}\orcidID{0000-0002-7808-6567} \and
Miguel A. Ferrer\inst{2}\orcidID{0000-0003-4913-4010} \and
Andreas Fischer\inst{3}\orcidID{0000-0003-0069-3436} \and
Josep Lladós\inst{1}\orcidID{0000-0002-4533-4739} \and
Cristina Martín\and
Eloy Opisso\inst{4}\orcidID{0000-0002-6868-6737} \and
Réjean Plamondon\inst{5}\orcidID{0000-0002-4903-7539} \and
Anna Scius-Bertrand\inst{3} \and
Josep Maria Tormos\inst{4}\orcidID{0000-0002-8764-2289} 
}
\authorrunning{A. Fornés et al.}
%
\institute{Computer Vision Center, Computer Science Department, \\Universitat Aut\`{o}noma de Barcelona, Spain\\
\email{\{afornes, abensalah, jchen, josep\}@cvc.uab.es}\\
 \and
Universidad de Las Palmas de Gran Canaria, Spain\\
\email{\{cristina.carmona, miguelangel.ferrer\}@ulpgc.es}\\
 \and
Institute of Complex Systems, University of Applied Sciences and Arts Western Switzerland, Fribourg, Switzerland\
\email{andreas.fischer@unifr.ch, Anna.Scius-Bertrand@hefr.ch}\\
 \and
Institut Guttmann, Neurorehabilitation Institute, Camí de Can s/n 08916 Badalona, Spain\
\email{\{cmartin, eopisso, jmtormos\}@guttmann.com}\\\
and
D\'{e}partement de G\'{e}nie \'{E}lectrique, Polytechnique Montr\'{e}al, Montr\'{e}al, Canada\ 
\email{rejean.plamondon@polymtl.ca}}
\maketitle              

\begin{abstract}
This project explores the feasibility of remote patient monitoring based on the analysis of 3D movements captured with smartwatches. We base our analysis on the Kinematic Theory of Rapid Human Movement. We have validated our research in a real case scenario for stroke rehabilitation at the Guttmann Institute\footnote{https://www.guttmann.com/en/} (neurorehabilitation hospital), showing promising results. Our work could have a great impact in remote healthcare applications, improving the medical efficiency and reducing the healthcare costs. Future steps include more clinical validation, developing multi-modal analysis architectures (analysing data from sensors, images, audio, etc.), and exploring the application of our technology to monitor other neurodegenerative diseases.

\end{abstract}

\section{Introduction}

Stroke, defined as the lack of blood flow or bleeding in the brain~\cite{stroke_def}, is the second leading cause of death in Europe. Moreover, experts estimate that strokes will rise dramatically in the next 20 years due to an ageing population~\footnote{The Burden of Stroke in Europe: \url{http://www.strokeeurope.eu/}}. Moreover, 60\% of the survivors have different degrees of disability, with a socio-economic impact of the first magnitude for the patient~\cite{economic}~\cite{economic1}, their environment, the health system and the society in general~\cite{depression}~\cite{depression1}. Therefore, in addition to stroke prevention, it is crucial to find personalized and suitable treatments during stroke rehabilitation, the most important phase of stroke survivors.

The Kinematic Theory of Rapid Human Movement~\cite{Plamondon1}~\cite{Plamondon2}~\cite{Plamondon3} provides a mathematical description of the movements made by individuals, reflecting the behaviour of their neuromuscular system. 
It has demonstrated a great potential for analysing fingers, hand, eye, head, trunk and arm movements as well as speech. According to the lognormal principle, the motor learning process and its deterioration with aging can be followed, allowing to monitor neuromuscular diseases in terms of the alteration of the ideal parameters. O’Reilly et al.\cite{o2014linking} showed that brain stroke risk factors can be associated with the deterioration of many cognitive and psychomotor characteristics. The psychomotor tests demonstrated that the features extracted from the kinematic motion analysis of handwriting were successfully correlated with risk factors (e.g. obesity, diabetes, hypertension, etc.). 

However, the use of the Kinematic Theory in monitoring rehabilitation processes is a challenge: it requires to collect and to analyse the movement data using robust, efficient and task oriented lognormal parameter extraction algorithms. These constraints must be removed to develop a universal tool for brain stroke treatments and rehabilitation. Stroke patients, especially in early stages of the recovery treatment, cannot write using a stylus on a tablet device, so most of the analysis of their motor skills improvement is based on simple hands or arms movements. 

Recently, inertial and magnetic sensors, including accelerometers, gyroscopes and magnetometers, have been incorporated into wearables, such as smartbands, to assess, among others, the biomechanics of sports performance. These devices are increasingly popular, which make us propose the hand/arm movements as a source to extract the lognormal patterns. Moreover, these devices are not intrusive, so they could be used for continuous remote patient monitoring (RPM) in the rehabilitation stages and during the routine daily life of patients, improving the medical efficiency and reducing the healthcare costs.

For the above mentioned reasons, we aim to explore the use of the Kinematic Theory of Rapid Human Movements for analysing continuous 3D movements captured with smartwatches (a worldwide affordable and non-intrusive technology), and thus, to provide an objective estimator of the improvement of the patients’ motor abilities in stroke rehabilitation. 

This paper describes the RPM3D project \footnote{\url{http://dag.cvc.uab.es/patientmonitoring/}} \cite{fornes2020exploring}, which aims to make a step forward towards the removal of such constraints to develop a universal tool for monitoring rehabilitation processes. Indeed, such a tool can have a great impact in remote health care tasks in general. The integration of an analytic tool in a consumer and affordable technology such as smartwatches (instead of high-end clinical devices) could be used for continuous remote patient monitoring in the rehabilitation stages of different neuromuscular diseases, improving the medical efficiency and reducing the healthcare costs. 

\begin{figure}
\includegraphics[width=\textwidth]{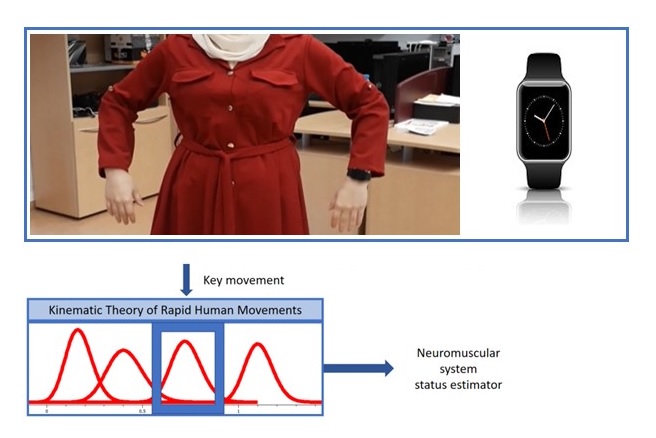}
\caption{Overview of the pipeline.} \label{pipeline}
\end{figure}

The overview of our approach is shown in Figure~\ref{pipeline}.
The main project results are the following: 
\begin{itemize}
    \item We have developed a smartwatch application to record data from the inertial sensors of smartwatches (concretely, the Apple Watch). 
    \item We have proposed a model to segment and classify the relevant gestures in continuous 3D movements for their posterior analysis. 
    \item We have adapted the parameter extraction algorithms of the kinematic model to these relevant 3D movements captured with the smartwatch. 
    \item We have defined the experimental protocol and validated our research in a real case scenario for stroke rehabilitation at the Guttmann Institute (neurorehabilitation hospital).
\end{itemize}
	
The innovation potential of this project is the provision of a new tool to obtain significant measures of the human movement of patients of brain strokes in the rehabilitation phase using wearable devices such as smartwatches. Conveniently calibrated, this tool can be seen as a \emph{thermometer} of the human neuromotor system, and with the appropriate interpretation (according to the correlation with the clinical indicators), medical doctors will be able to make decisions on the rehabilitation prescription and treatment of patients. 

The rest of the paper is organized as follows. In Section 2, we overview the state of the art. Next, in Section 3, we describe the application protocol and the capturing of data from the smartwatches. Section 4 is devoted to the recognition of movements, whereas Section 5 describes the kinematic analysis performed. Section 6 is devoted to the conclusions and future work.

\section{State of the Art}

Assessing the physical condition in rehabilitation scenarios is challenging because it involves Human Activity Recognition (HAR)~\cite{har} and kinematic analysis. 

HAR methods must deal with intraclass variability and interclass similarities~\cite{intraclass}~\cite{imbalance}. Also, the detection of target (relevant) movements is difficult due to the diversity of non-target movements. In continuous time series data, the challenge is to detect and segment those subsequences (target movements) so that they can be properly analysed by the kinematic model. This is especially difficult when the movements are non-repetitive and that is why a major part of activity recognition works deal only with repetitive(periodic) movements such as: walking\cite{walking}, stair ascent or descent~\cite{stairs}, running, sport exercises~\cite{sport}...

HAR is about seeking high-level knowledge that describes human activities, ergo HAR benefited broadly from deep learning since this latter one can provide automatic feature extraction~\cite{review1}~\cite{review2}~\cite{review3}.

At the same time, traditional machine learning like Support Vector Machines (SVMs)~\cite{svm}~\cite{svm1}, K-Nearest-Neighbours (KNNs)~\cite{knn}~\cite{knn1} still provide an efficient accurate solution for HAR tasks due to the fact that they perform better in few data problems which is the case of most HAR tasks that suffer from data scarce.

As mentioned in the introduction, the Kinematic Theory of Rapid Human Movement ~\cite{lognormal} has demonstrated a great potential for monitoring neuromuscular diseases, but it requires robust algorithms to estimate the model parameters with an excellent precision for a meaningful neuromuscular analysis. So far, most algorithms (Idelog \cite{Idelog} and Robust XZERO \cite{xZero1,Xzero2}) have mainly focused on 1D and 2D movements in a controlled scenario, e.g. pen movements on a tablet computer . This constraint makes the approach unrealistic for stroke rehabilitation. Stroke patients have severe mobility limitations, especially in early stages, so the analysis of their motor skills improvement is based on simple hands or arms movements. Thus, the recently proposed 3D algorithm~\cite{extension} must be adapted to continuous movements in unconstrained scenarios (closer to real use cases). Finally, the hardware is an extra difficulty, because the smartwatch could be less accurate than clinical devices.

In summary, the challenges are the following:
\begin{itemize}
    \item The use of sensors from consumer devices instead of clinical devices, which can decrease the quality of the data for the application of the kinematic model. 
    \item The extraction of the model parameters from the continuous 3D movement sequences for their posterior analysis. 
    \item The accurate detection, segmentation and analysis of the target movements in uncontrolled scenarios.
\end{itemize}

\section{Application Protocol and Data Capturing}

Next, we describe the application protocol and the recorded movements.

\subsection{Application Protocol}
\begin{figure}[t]
  \centering
  \begin{tabular}{cccc}
    \includegraphics[width=0.23\textwidth, height=4cm]{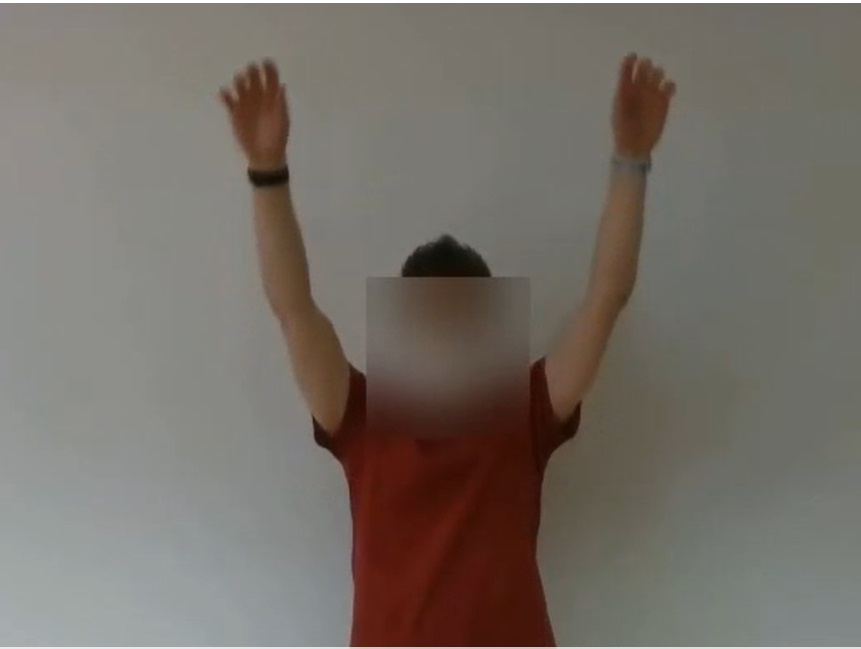} &
    \includegraphics[width=0.23\textwidth, height=4cm]{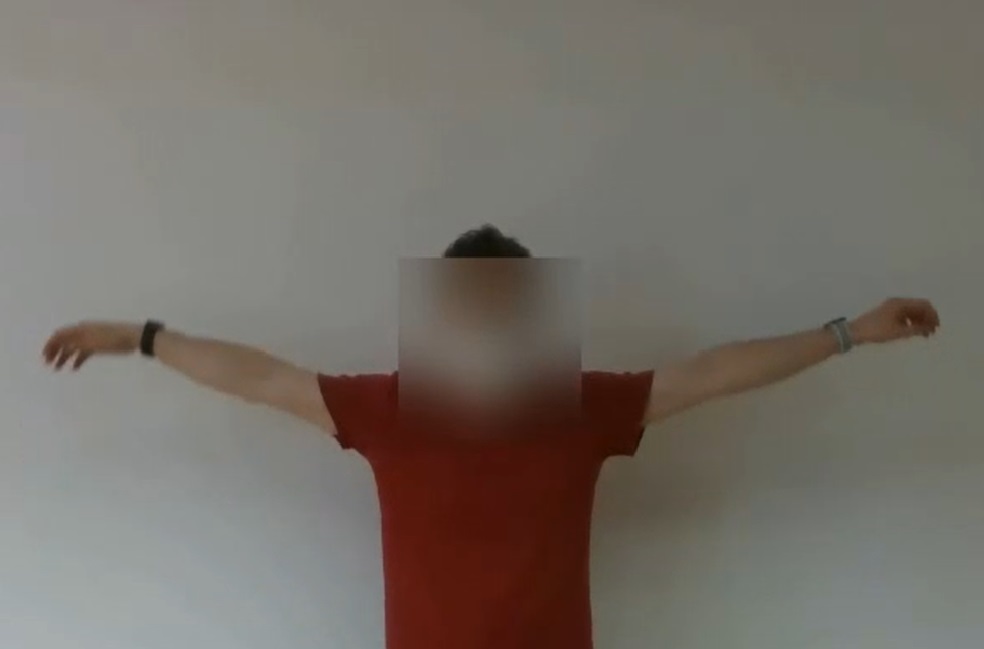} &
    \includegraphics[width=0.23\textwidth,height=4cm]{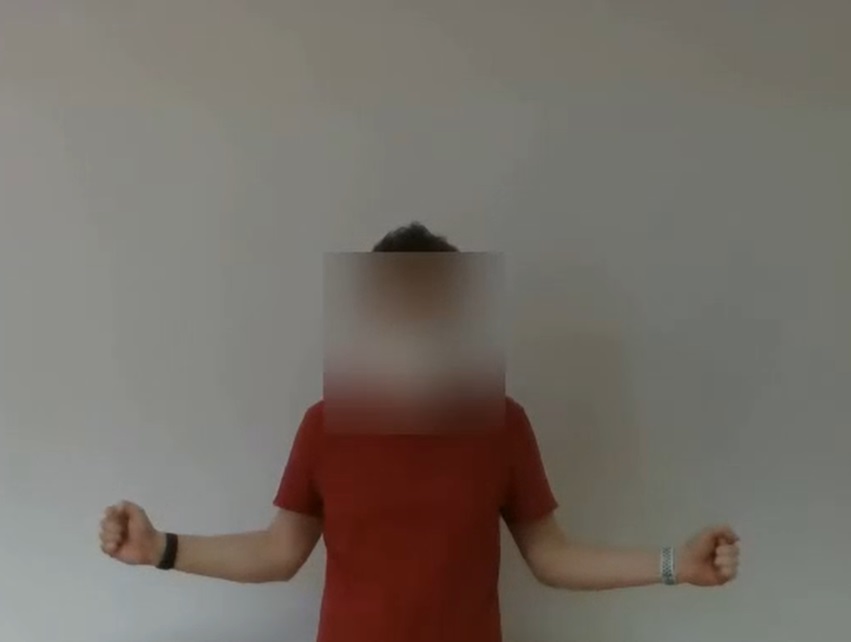} &
    \includegraphics[width=0.23\textwidth,height=4cm]{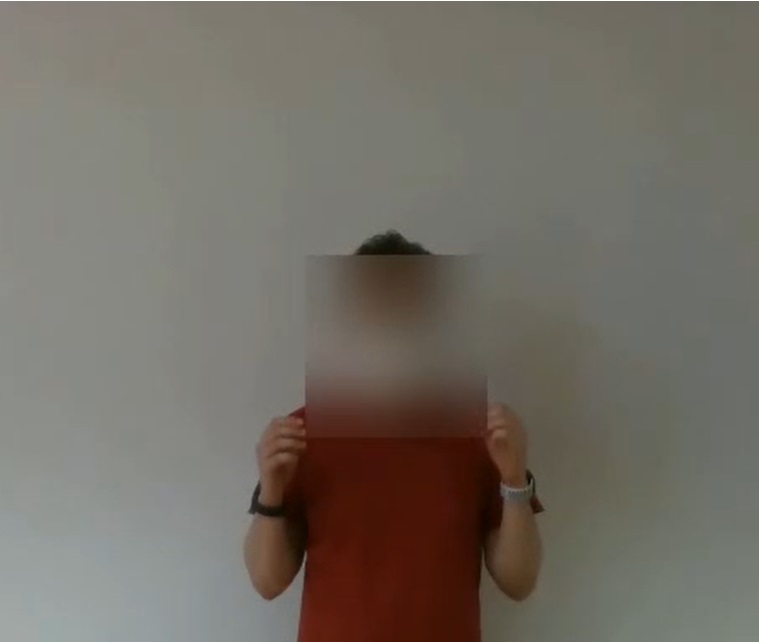} \\
  a) & b) &  c) & d) \\
  \end{tabular}\\
  \caption{Target movements. a) Movement 1; b) Movement 2; c) Movement 3; d) Movement 4. }
    \label{fig:movements}
\end{figure}

We have designed an upper-limb assessment pipeline inspired by the Fugl-Meyer Assessment scale, an index to assess the sensorimotor impairment in stroke patients. Concretely, we have defined four target (non-repetitive) movements (see Figure~\ref{fig:movements}), based on the following joint movements: 
\begin{enumerate}
    \item Shoulder extension/flexion
    \item Shoulder adduction/abduction
    \item External/internal shoulder rotation
    \item Elbow flexion/extension
\end{enumerate}

We have recorded these movements in two scenarios: 
\begin{itemize}
    \item L1 is a constrained scenario which consists in performing the same target movement in a sequence, but alternating the arm (left, right or both). 
    \item L2 is an unconstrained scenario, where target movements appear inside longer sequences that include non-target movements (e.g. common daily life activities like eating, pouring water into a glass, brushing your teeth, scratching the ear, etc.). 
\end{itemize}

As a proof of concept, we have recorded data from 25 healthy individuals and 4 patients from Guttmann Institute.
Out of the 25 healthy individuals, 48\% are women and 52\% are men. While for the patient population, there is  one woman and 3 men. 
Healthy and patient individuals' age range between 20 and 60 years.

The users wear two watches, one in each wrist. Patients data was recorded along four sessions with an interval of one to two weeks, while healthy individuals' data was recorded in one session.

\subsection{Data Capturing}
We have developed an application for the Apple Watch 4 to record the sequences of movements, as shown in Fig\ref{pipeline}.
The user-generated acceleration (without gravity) for all three axes of the device, unbiased gyroscope (rotation rate), magnetometer, altitude (Euler angles) and temporal information data have been recorded in the watch’s internal memory at 100Hz sampling rate. 

The two watches are synchronised thanks to an audio signal.  
Afterwards, the data is transmitted to the mobile phone and the cloud service. Finally, the signal is {\bfseries preprocessed} to minimize the sensor drift, which often leads to inaccurate measures and larger accumulated error.
\section{Human Activity Recognition}
We have used the Euler angles and the linear acceleration. To detect the target movements in the unconstrained scenario L2, we explored two segmentation options: 
\begin{enumerate}
\item Segmenting the complete sequence using non-overlapping sliding windows (namely action recognition). 
\item Picking the positive peaks in the signal as candidates to be relevant movements (namely gesture spotting). 
\end{enumerate}
We have also explored two classification methods. First, SVMs, a machine learning approach typically used in HAR, together with the following feature vector set: the mean, the minimum, the maximum and the standard variation of the window. Second, Convolutional Neural Networks (CNN), a deep learning model in which the input is the linear acceleration signal instead of a feature vector set. More details can be found at \cite{icprAsma2021}.

\begin{table}[htbp]
  \centering
  \caption{HAR classification and spotting results}
   \resizebox{\columnwidth}{!}{
    \begin{tabular}{|p{5.135em}|c|c|c|c|c|c|}
\cline{2-7}    \multicolumn{1}{r|}{} & \multicolumn{4}{p{18.86em}|}{\textbf{Healthy Individuals}} & \multicolumn{2}{p{11.41em}|}{\textbf{Patients}} \bigstrut\\
\cline{2-7}    \multicolumn{1}{r|}{} & \multicolumn{2}{p{6.275em}|}{\textbf{Action Recognition}} & \multicolumn{2}{p{6.275em}|}{\textbf{Gesture Spotting}} & \multicolumn{1}{p{6.275em}|}{\textbf{Action Recognition}} & \multicolumn{1}{p{6.275em}|}{\textbf{Gesture Spotting}} \bigstrut\\
    \hline
    \textbf{Scenario} & \multicolumn{1}{p{5.135em}|}{\textbf{SVM}} & \multicolumn{1}{p{5.135em}|}{\textbf{CNN}} & \multicolumn{1}{p{5.135em}|}{\textbf{SVM}} & \multicolumn{1}{p{5.135em}|}{\textbf{CNN}} & \multicolumn{1}{p{5.135em}|}{\textbf{SVM}} & \multicolumn{1}{p{5.135em}|}{\textbf{SVM}} \bigstrut\\
    \hline
    \textit{\textbf{L1}} & \textbf{84\%} & \textbf{65\%} & \textbf{55\%} & \textbf{60\%} & \textbf{56\%} & \textbf{41\%} \bigstrut\\
    \hline
    \textit{\textbf{L2}} & \textbf{61\%} & \textbf{59\%} & \textbf{51\%} & \textbf{53\%} & \textbf{41\%} & \textbf{35\%} \bigstrut\\
    \hline
    \end{tabular}
    }%
  \label{tab:har_results}%
\end{table}%

As shown in Table~\ref{tab:har_results}, action recognition is preferable. In healthy individuals, the SVM classifier obtains better results (84\% in L1 and 61\% in L2) than the CNN one (65\% in L1 and 59\% in L2) because the CNN is a data hungry method. Concerning gesture classification, the results by the two classifiers are similar. In patients, the accuracy in the unhealthy body part decreases (56\% in L1 and 41\% in L2) in comparison with their healthy side (84,5\% in L1 and 61\% in L2), because these movements are less accurate due to their loss of motor function.

\section{Kinematic Analysis}
The Kinematic Theory of Rapid Human Movements describes the resulting speed of a neuromuscular system action as a lognormal function~\cite{Plamondon1}~\cite{Plamondon2}~\cite{Plamondon3}. To analyse the 3D movements captured by smartwatches, we utilize a recently proposed 3D extension of the Sigma-Lognormal model~\cite{extension} to decompose observed 3D movements into sequences of elementary movements with lognormal speed. There are several model parameters that can be analysed with a view to the patients’ motor abilities.

Here, we focus on the signal-to-noise-ratio (SNR) between the observed trajectory of the smartwatch and the reconstructed trajectory using the analytical model. A high SNR indicates a high model quality, i.e. a good representation of the 3D movement. Furthermore, healthy subjects tend to achieve a higher SNR than patients with motor control problems~\cite{lognormal}. 

\begin{table}[htbp]
  \centering
  \caption{Kinematic analysis mean standard deviation}
    \begin{tabular}{|c|c|c|}
\cline{2-3}    \multicolumn{1}{c|}{} & Healthy Individuals & Patients \bigstrut\\
    \hline
    Samples & 649   & 126 \bigstrut\\
    \hline
    Duration [s] & 4.1 ± 1.0 & 4.9 ± 0.8 \bigstrut\\
    \hline
    Number of Lognormals & 17.3 ± 4.7 & 17.6 ± 4.5 \bigstrut\\
    \hline
    SNR [dB] & 22.2 ± 2.8 & 21.3 ± 2.1 \bigstrut\\
    \hline
    \end{tabular}%
  \label{tab:kin_results}%
\end{table}%

\begin{figure}[h]
  \centering
  \includegraphics[width=0.8\textwidth]{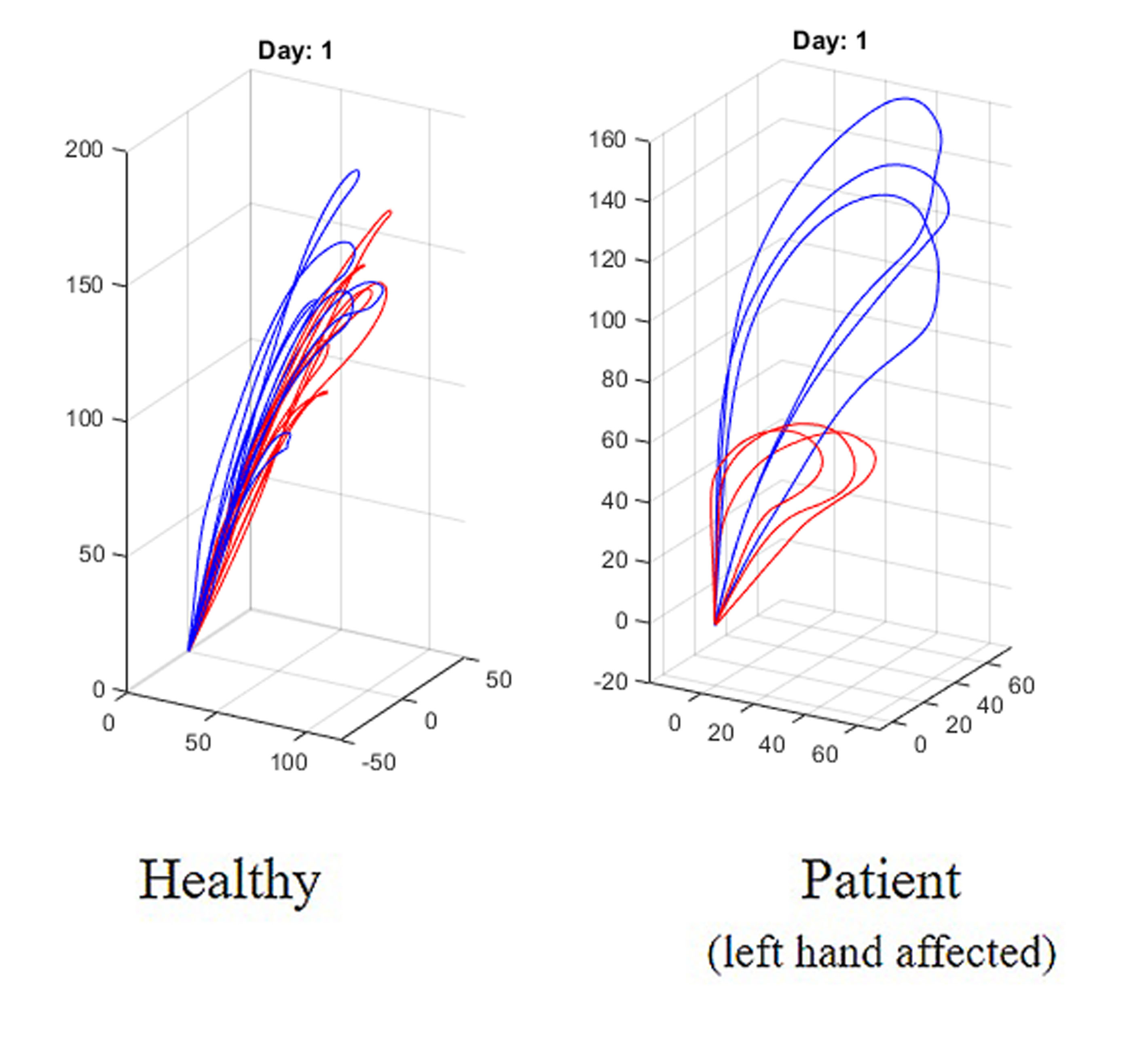}
\caption{Kinematic analysis results.} \label{kin_results}
\end{figure}

Table~\ref{tab:kin_results} and Fig~\ref{kin_results} present the first results of our kinematic analysis, comparing 649 movements from 25 healthy individuals with 126 movements from 4 patients.
In both cases an excellent SNR is achieved, indicating that the 3D Sigma-Lognormal model is suitable for analysing the smartwatch movements. Furthermore, we observe that the patients needed more time to execute the movements, more lognormals were needed to model the patients’ movements, and a lower SNR was achieved. The difference in SNR is statistically significant (Mann-Whitney U test, p < 0.0001). These observations are consistent with the lognormality principle~\cite{extension} and encourage a more detailed kinematic analysis of the patients’ motor abilities based on the Kinematic Theory.


\section{Conclusion and Future Steps}

In this paper, we have presented the RPM3D project, which aims to ease the monitoring of patients during the neurorehabilitation stages.

In the future, we plan to focus on the continuous and remote monitoring of the patients’ neuromotor status. Concretely:

\begin{itemize}
    \item We will to perform more clinical validation through an exhaustive analysis of the correspondence between the kinematic analysis and the clinicians’ estimations. We will also continue the comparative analysis between healthy users and patients.
    \item We will explore the use of other lower-cost wearables (e.g. smarbands) and also, the possibility to combine the sensor data with video images or speech. Also, we would like to recognize functional (purposeful) movements to determine the degree of integration of the affected side of the body in the patients’ daily life actions.
    \item We will explore the adaptation of our approach for monitoring patients suffering from Multiple Sclerosis or Parkinson diseases, the ageing effects in elderly people, the effects of medication in clinical trials, etc.
\end{itemize}

\section{Acknowledgement}
This work has been partially supported by the Spanish project RTI2018-095645-B-C21, the CERCA Program / Generalitat de Catalunya and the FI fellowship AGAUR 2020 FI-SDUR 00497 (with the support of the Secretaria d’Universitats i Recerca of the Generalitat de Catalunya and the Fons Social Europeu).

%
%
%
%

\bibliographystyle{unsrt}
\bibliography{bib.bib}

\begin{thebibliography}{10}

\bibitem{stroke_def}
Alexander~P Coupland, Ankur Thapar, Mahim~I Qureshi, Harri Jenkins, and Alun~H
  Davies.
\newblock The definition of stroke.
\newblock {\em Journal of the Royal Society of Medicine}, 110(1):9--12, 2017.
\newblock Query date: 2022-03-24 13:39:45.

\bibitem{economic}
J.~Majersik and D.~Woo.
\newblock The enormous financial impact of stroke disability, 2020.
\newblock 2 cites:.

\bibitem{economic1}
S.~Rajsic, H.~Gothe, H.~Borba, G.~Sroczynski, J.~Vujičić, T.~Toell, and
  U.~Siebert.
\newblock Economic burden of stroke: a systematic review on post-stroke care,
  2018.
\newblock Query date: 2022-03-24 10:29:44.

\bibitem{depression}
Francesco Bartoli, Carmen Di~Brita, Cristina Crocamo, Massimo Clerici, and
  Giuseppe Carrà.
\newblock Early post-stroke depression and mortality: Meta-analysis and
  meta-regression.
\newblock {\em Frontiers in Psychiatry}, 9, 2018.

\bibitem{depression1}
A.~Hussein, I.~Idris, M.~Abbasher, H.~Abbashar, and K.~Mohamed~Ahmed Abbasher.
\newblock Post stroke depression.
\newblock {\em Journal of the Neurological Sciences}, 405:70, 2019.
\newblock Abstracts from the World Congress of Neurology (WCN 2019).

\bibitem{Plamondon1}
R\'{e}jean Plamondon.
\newblock A kinematic theory of rapid human movements.
\newblock {\em Biol. Cybern.}, 72(4):309–320, mar 1995.

\bibitem{Plamondon2}
R.~Plamondon.
\newblock {A Kinematic Theory of Rapid Human Movements - Part II. Movement time
  and control}.
\newblock {\em Biological Cybernetics}, 72(4):309--320, 1995.

\bibitem{Plamondon3}
R.~Plamondon.
\newblock {A kinematic theory of rapid human movements: Part III. Kinetic
  outcomes}.
\newblock {\em Biological Cybernetics}, 78(2):133--145, 1998.

\bibitem{o2014linking}
Christian O'Reilly, R{\'e}jean Plamondon, and Louise-H{\'e}l{\`e}ne Lebrun.
\newblock Linking brain stroke risk factors to human movement features for the
  development of preventive tools.
\newblock {\em Frontiers in aging neuroscience}, 6:150, 2014.

\bibitem{fornes2020exploring}
Alicia Forn{\'e}s, Asma Bensalah, Mar{\'\i}a~Cristina Carmona~Duarte,
  Miguel~{\'A}ngel Ferrer~Ballester, Andreas Fischer, Josep Llad{\'o}s, Eloy
  Opisso, R{\'e}jean Plamondon, and Josep~Maria Tormos.
\newblock Exploring the 3d kinematics for brain stroke rehabilitation.
\newblock In Réjean Plamondon, Angelo Marcelli, and Miguel Ángel Ferrer,
  editors, {\em The Lognormality Principle and its Applications in e-Security,
  e-Learning and e-Health}, pages 349--352. World Scientific Publishing, 2020.

\bibitem{har}
Upal Mahbub and Md~Atiqur~Rahman Ahad.
\newblock Advances in human action, activity and gesture recognition.
\newblock {\em Pattern Recognition Letters}, 155:186--190, 2022.
\newblock Query date: 2022-03-23 14:26:15.

\bibitem{intraclass}
Akila K and Chitrakala S.
\newblock An efficient method to resolve intraclass variability using highly
  refined hog description model for human action recognition.
\newblock {\em Concurrency and Computation: Practice and Experience}, 31(12),
  2018.
\newblock Query date: 2022-03-23 14:31:14.

\bibitem{imbalance}
Fayez Alharbi, Lahcen Ouarbya, and Jamie~A Ward.
\newblock Comparing sampling strategies for tackling imbalanced data in human
  activity recognition.
\newblock {\em Sensors}, 22(4):1373--1373, 2022.
\newblock Query date: 2022-03-23 14:29:13.

\bibitem{walking}
Yashi Nan, Nigel Lovell, Kejia Wang, Kim Delbaere, and Kim van Schooten.
\newblock Deep learning for activity recognition in older people using a
  pocket-worn smartphone.
\newblock {\em Sensors}, 20:7195, 12 2020.

\bibitem{stairs}
Vijay~Bhaskar Semwal, Anjali Gupta, and Praveen Lalwani.
\newblock An optimized hybrid deep learning model using ensemble learning
  approach for human walking activities recognition.
\newblock {\em Journal Of Supercomputing}, 77(11):12256--12279, 2021.
\newblock 10 cites:.

\bibitem{sport}
Jenny Margarito, Rim Helaoui, Anna~M. Bianchi, Francesco Sartor, and Alberto~G.
  Bonomi.
\newblock User-independent recognition of sports activities from a single
  wrist-worn accelerometer: A template-matching-based approach.
\newblock {\em IEEE Transactions on Biomedical Engineering}, 63(4):788--796,
  2016.

\bibitem{review1}
M.~Straczkiewicz, Peter James, and J.~Onnela.
\newblock A systematic review of smartphone-based human activity recognition
  methods for health research, 2021.
\newblock 6 cites:.

\bibitem{review2}
Netzahualcóyotl Hernández, J.~Lundström, J.~Favela, I.~McChesney, and
  B.~Arnrich.
\newblock Literature review on transfer learning for human activity recognition
  using mobile and wearable devices with environmental technology, 2020.
\newblock 21 cites:.

\bibitem{review3}
Lifang Wu, Qi~Wang, Meng Jian, Y.~Qiao, and Boxuan Zhao.
\newblock A comprehensive review of group activity recognition in videos, 2021.
\newblock 8 cites:.

\bibitem{svm}
Zhenghua Chen, Qingchang Zhu, Yeng~Chai Soh, and Le~Zhang.
\newblock Robust human activity recognition using smartphone sensors via ct-pca
  and online svm.
\newblock {\em IEEE Transactions on Industrial Informatics}, 13:3070--3080,
  2017.

\bibitem{svm1}
K.~G.~Manosha Chathuramali and Ranga Rodrigo.
\newblock Faster human activity recognition with svm.
\newblock {\em International Conference on Advances in ICT for Emerging Regions
  (ICTer2012)}, pages 197--203, 2012.

\bibitem{knn}
Zongying Liu, Shaoxi Li, Jiangling Hao, Jingfeng Hu, and Mingyang Pan.
\newblock An efficient and fast model reduced kernel knn for human activity
  recognition.
\newblock {\em Journal of Advanced Transportation}, 2021:1--9, 2021.

\bibitem{knn1}
Paulo J.~S. Ferreira, Jo{\~a}o~MP Cardoso, and Jo{\~a}o Mendes-Moreira.
\newblock knn prototyping schemes for embedded human activity recognition with
  online learning.
\newblock {\em Comput.}, 9:96, 2020.

\bibitem{lognormal}
Réjean Plamondon, Christian O'Reilly, Céline Rémi, and Thérésa Duval.
\newblock The lognormal handwriter: learning, performing, and declining.
\newblock {\em Frontiers in Psychology}, 4, 2013.

\bibitem{Idelog}
Miguel~A. Ferrer, Moises Diaz, Cristina Carmona-Duarte, and Rejean Plamondon.
\newblock {IDeLog: Iterative Dual Spatial and Kinematic Extraction of
  Sigma-Lognormal Parameters}.
\newblock {\em IEEE Transactions on Pattern Analysis and Machine Intelligence},
  42(1):114--125, 2020.

\bibitem{xZero1}
Christian O’Reilly and Réjean Plamondon.
\newblock Development of a sigma–lognormal representation for on-line
  signatures.
\newblock {\em Pattern Recognition}, 42(12):3324--3337, 2009.
\newblock New Frontiers in Handwriting Recognition.

\bibitem{Xzero2}
Moussa Djioua and R{\'e}jean Plamondon.
\newblock A new algorithm and system for the extraction of delta-lognormal
  parameters.
\newblock 2008.

\bibitem{extension}
Andreas Fischer, Roman Schindler, Manuel Bouillon, and Réjean Plamondon.
\newblock {\em Modeling 3D Movements with the Kinematic Theory of Rapid Human
  Movements}, chapter Chapter 15, pages 327--342.

\bibitem{icprAsma2021}
Asma Bensalah, Jialuo Chen, Alicia Forn{\'e}s, Cristina Carmona-Duarte, Josep
  Llad{\'o}s, and Miguel~{\'A}ngel Ferrer.
\newblock Towards stroke patients' upper-limb automatic motor assessment using
  smartwatches.
\newblock In {\em International Workshop on Artificial Intelligence for
  Healthcare Applications (IAHA). ICPR Workshops}, pages 476--489. Springer
  International Publishing, 2021.

\end{thebibliography}
\end{document}